\documentstyle[pra,aps,epsf,twocolumn,array]{revtex}
\begin{document}
\draft

\title{Entangling capacity of global phases and implications for Deutsch-Jozsa algorithm}

\author{H.~Azuma$^{(1)}$\thanks{On leave from Canon Research Center, 5-1,
Morinosato-Wakamiya, Atsugi-shi, Kanagawa, 243-0193, Japan,}\thanks{hiroo.azuma@qubit.org},
S.~Bose$^{(1)}$\thanks{sougato.bose@qubit.org}
and V.~Vedral$^{(2)}$\thanks{vlatko.vedral@qubit.org}}

\address{$^{(1)}$Centre for Quantum Computation,
Clarendon Laboratory, Parks Road, Oxford OX1 3PU,UK \\
$^{(2)}$Optics Section, Blackett Laboratory, Imperial College,
London SW7 2BZ, UK}

\date{\today}

\maketitle

\begin{abstract}
We investigate the creation of entanglement by the application of
phases whose value depends on the state of a collection of qubits.
First we give the necessary and sufficient conditions for a given
set of phases to result in the creation of entanglement in a
state comprising of an arbitrary number of qubits. Then we analyze
the creation of entanglement between any two qubits in three-qubit
pure and mixed states. We use our result to prove that
entanglement is necessary for Deutsch-Jozsa algorithm to have an
exponential advantage over its classical counterpart.
\end{abstract}



\section{Introduction}
In recent years, entanglement has become an important resource for
quantum communications \cite{bennett00}. Quantum computation
\cite{Deu85}, which is more efficient than classical computation
for certain problems \cite{dj,Shor,Grov}, could also potentially
owe its efficiency to entanglement \cite{jozsa,ej,lindpop}. Though
the precise role of entanglement in quantum computation is not yet
well understood, entangled states are certainly generated during
the course of certain quantum computations. A quantum computation,
when halted at an appropriate point, can be regarded as a method
of generating entanglement. Typically, a quantum computation is a
multiparticle interference experiment with different phases
applied to distinct multiparticle states \cite{mosca}. In general,
the phases applied to the multiparticle states during a quantum
computation are {\em global phases} as they depend on the total
state of a collection of qubits. In this paper, we will
investigate the types of entanglement generated by such global
phases and the conditions under which such phases do not generate
any entanglement.

\begin{figure}[htbp]
\begin{center}
\leavevmode
\epsfxsize=8cm
\epsfbox{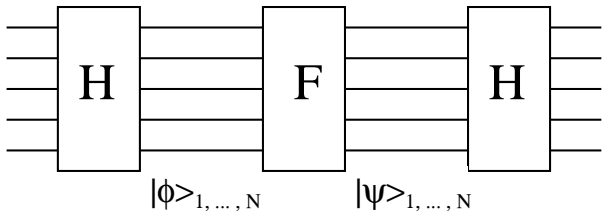}
\caption{\narrowtext A typical quantum computation network.}
\label{fig1}
\end{center}
\end{figure}

The model of quantum computation which motivates our work is that
presented by Cleve, Ekert, Macchiavello and Mosca \cite{mosca}.
This model (with a slight alteration which does not change its
principal ingredient) is illustrated in Fig.~\ref{fig1}. Each of
the qubits, initially in the $|0\rangle$ state, is first
transformed according to a Hadamard transformation. This is shown
in the figure by the giant Hadamard transformation acting on all
the qubits and converts the total state of the qubits to
\begin{equation}
|\phi\rangle_{1,\cdots,N}=\frac{1}{2^{N/2}} \sum_{j=0}^{2^{N}-1}
|j\rangle,
\end{equation}
where $N$ is the total number of qubits and the index $j$ labels
the $2^N$ possible states of the type $|j_1, ... ,j_N\rangle$ in
which each $j_i=0$ or $1$. $|\phi\rangle_{1,\cdots,N}$ is a
disentangled state. A state dependent global phase $f(j)$ is now
applied to each state $|j\rangle$. This is shown as the second
giant transformation $F$ in the figure. This converts the total
state to
\begin{equation}
|\psi\rangle_{1,\cdots,N}
=
\frac{1}{2^{N/2}}
\sum_{j=0}^{2^{N}-1}
e^{if(j)}|j\rangle,
\label{N-qubit-phase-eq}
\end{equation}
where $\{f(j)\}$ are real and $0\leq f(j)< 2\pi$ ($f(j)=2\pi$ is
reassigned the value $0$). This state $|\psi\rangle_{1,\cdots,N}$,
generated as a result of global phases, can be entangled.  We
propose to halt the quantum computation at this stage and
investigate the amount of entanglement generated. A complete
quantum computation, of course, consists of one more step in
which another giant Hadamard transformation is applied to all the
qubits as shown in Fig.~\ref{fig1}. But in this paper we are interested in
the entanglement of the state {\em prior} to this last
transformation.

The entanglement of $|\psi\rangle_{1,\cdots,N}$
comes from the global phase factors $f(j)$. First, we study
conditions on the phase function $f(j)$ for the state
$|\psi\rangle_{1,\cdots,N}$ to be disentangled. Next, we derive
the entanglement of three-qubit pure states ($N=3$) for the special
case in which only one or two of the global phase parameters are
nonzero. We study variation of the entanglement as a function of
one global phase parameter for a mixed state of three qubits by
numerical calculations. Finally, we discuss the implications of
this type of entanglement arising in Deutsch-Jozsa algorithm. In
particular we show that for obtaining exponential advantage over
its classical counterpart, entangled states must necessarily
arise in Deutsch-Jozsa algorithm.

\section{Necessary and sufficient conditions
for the generation of entanglement by global phases}
\label{Disentanglement}

We first derive the conditions on $\{f(j)\}$ for
$|\psi\rangle_{1,\cdots,N}$ to be disentangled, i.e.,
\begin{equation}
|\psi\rangle_{1,\cdots,N}
=|\psi_{1}\rangle\otimes\cdots\otimes|\psi_{N}\rangle.
\end{equation}
In a case of two qubits ($N=2$),
we can write the condition as follows,
\begin{equation}
[f(0)-f(1)]-[f(2)-f(3)]=2\pi n,
\end{equation}
where $n$ is an arbitrary integer.

We now consider
the case of three qubits,
\begin{equation}
|\psi_{ABC}\rangle
=
\frac{1}{2^{3/2}}
\sum_{j=0}^{7}
e^{if(j)}|j\rangle.
\label{3-qubit-phase-eq}
\end{equation}
First, we derive condition
that the qubit $C$ is disentangled from the qubits $AB$.
The density matrix of the qubit $C$ is given by
\begin{equation}
\rho_{C}
=\mbox{tr}_{AB}(|\psi_{ABC}\rangle\langle\psi_{ABC}|)
=\frac{1}{8}
\left (
\begin{array}{cc}
4 & \gamma \\
\gamma^{*} & 4\\
\end{array}
\right ),
\end{equation}
with taking a basis $\{|0\rangle, |1\rangle\}$
and
\begin{eqnarray}
\gamma
&=&
e^{i[f(0)-f(1)]}+e^{i[f(2)-f(3)]} \nonumber \\
&&
+e^{i[f(4)-f(5)]}+e^{i[f(6)-f(7)]}.
\end{eqnarray}
(From now on,
when we give a matrix representation of
a density operator on a $2^{N}$-dimensional space,
we always take a logical basis of
$\{|x\rangle: x\in\{0,1\}^{N}\}$.)
If and only if $\mbox{tr}({\rho_{C}}^{2})=1$,
the qubit $C$ is disentangled from the qubits $AB$.
Hence we obtain the following constraints,
\begin{eqnarray}
{[}f(0)-f(1){]}-{[}f(2)-f(3){]}&=&2\pi n_{1},
\label{3-qubit-constraint1} \\
{[}f(0)-f(1){]}-{[}f(4)-f(5){]}&=&2\pi n_{2},
\label{3-qubit-constraint2} \\
{[}f(0)-f(1){]}-{[}f(6)-f(7){]}&=&2\pi n_{3}.
\label{3-qubit-constraint3}
\end{eqnarray}
Next, we consider the condition that the qubit $B$ is
disentangled from the qubits $AC$. From similar considerations
before, we obtain another constraint,
\begin{equation}
[f(0)-f(2)]-[f(4)-f(6)]=2\pi n_{4}.
\label{3-qubit-constraint4}
\end{equation}
From these results, we obtain four constraints,
Eqs.~(\ref{3-qubit-constraint1}), (\ref{3-qubit-constraint2}),
(\ref{3-qubit-constraint3}) and (\ref{3-qubit-constraint4}),
where $n_{1},\cdots,n_{4}$ are arbitrary integers, so that
$|\psi_{ABC}\rangle$ is disentangled perfectly.

Next consider the general case of $N$ qubits. Before deriving the condition
for $|\psi\rangle_{1,\cdots,N}$ to be disentangled, we think how
many constraints of $\{f(j)\}$ do we need to disentangled
$|\psi\rangle_{1,\cdots,N}$ completely. In
Eq.~(\ref{N-qubit-phase-eq}), the number of real parameters is
equal to $2^{N}$. On the other hand, if
$|\psi\rangle_{1,\cdots,N}$ is disentangled, we can describe it as
\begin{eqnarray}
\label{imp}
\lefteqn{|\psi\rangle_{1,\cdots,N}} \nonumber \\
&=&
e^{i\theta_{0}}
(|0\rangle+e^{i\theta_{1}}|1\rangle) \otimes\cdots\otimes
(|0\rangle+e^{i\theta_{N}}|1\rangle),
\end{eqnarray}
where $0\leq \theta_{i} < 2\pi$
for $0\leq \forall i \leq N$
and the number of real parameters is equal to $(N+1)$.
Therefore,
to disentangle $|\psi\rangle_{1,\cdots,N}$
to an $N$-qubit product state,
we need $[2^{N}-(N+1)]$ constraints.
The constraints are given as follows,
\begin{equation}
\begin{array}{l}
{[}f(0)-f(1){]}-{[}f(2)-f(3){]}=2\pi n_{1}, \\
{[}f(0)-f(1){]}-{[}f(4)-f(5){]}=2\pi n_{2}, \\
\quad\quad\vdots \\
{[}f(0)-f(1){]}-{[}f(2^{N}-2)-f(2^{N}-1){]} \\
\quad\quad\quad\quad\quad\quad =2\pi n_{2^{N-1}-1}, \\
{[}f(0)-f(2){]}-{[}f(4)-f(6){]}=2\pi m_{1}, \\
\quad\quad\vdots \\
{[}f(0)-f(2){]}-{[}f(2^{N}-4)-f(2^{N}-2){]} \\
\quad\quad\quad\quad\quad\quad =2\pi m_{2^{N-2}-1}, \\
\quad\quad\vdots \\
{[}f(0)-f(2^{N-2}){]}-{[}f(2^{N-1})-f(3\cdot 2^{N-2}){]} \\
\quad\quad\quad\quad\quad\quad =2\pi l,
\end{array}
\end{equation}
and we can confirm that the number of the above constraints is
\begin{equation}
\sum_{k=1}^{N}(2^{N-k}-1)=2^{N}-(N+1).
\end{equation}
As $|\psi\rangle_{1,\cdots,N}$ being disentangled automatically
implies that {\em all} the above constraints hold, if any of them
fail, $|\psi\rangle_{1,\cdots,N}$ is necessarily entangled. A
{\em sufficient} condition for global phase functions $f(j)$ to
produce entanglement is thus the violation of any of the above
constraints.

In a compact form the above
expression can be rewritten as
\begin{equation}
\label{eql2}
f(j)=\vec{\theta}\cdot\vec{j}+\theta_{0},
\quad
\mbox{(mod $2\pi$)}
\end{equation}
where $\vec{\theta}=(\theta_{1},\cdots,\theta_{N})$,
$\vec{j}=(j_1, ... ,j_N)$, and the components $j_i$ are
obtained from the binary expression of $j$ as $j_1, ... ,j_N$,
and `$\cdot$' means the inner product of $N$-component vectors. An
easy argument now proves that the violation of Eq.~(\ref{eql2})
is also a necessary condition for the generation of entanglement
by global phases. Consider a phase function $f(j)$ expressible in
the form of Eq.~(\ref{eql2}). Then the whole state, after
application of the global phases, can be rewritten in the form of
Eq.~(\ref{imp}). This is a disentangled state. This means that the
ability to express $f(j)$ in the form of Eq.~(\ref{eql2}) implies
no generation of entanglement. In other words, to generate
entanglement it is necessary to have a violation of
Eq.~(\ref{eql2}). We have thus found that the {\em necessary and
sufficient condition for the generation of entanglement by global
phases is the impossibility of the expansion given by
Eq.~(\ref{eql2}) of the global phase function}. In the subsequent
sections, we proceed to study the degree and type of entanglement
generated by some global phase functions which generate
entanglement.

\section{Entanglement between two qubits on three-qubit pure states}
\label{Entanglement-on-3qubit-pure-state}

The general problem of entanglement generation by global phase
functions for $N$ qubits is very complicated as it involves $2^{N}$
phase parameters. We will consider the simpler case of three-qubit
pure states that have just one or two nonzero phase parameters.
We first derive how the entanglement between two qubits of a
three-qubit pure state varies as a function of global phase
functions. For this, we evaluate the complete three-qubit pure
state after application of the global phases, compute the reduced
density matrix for any two qubits, and obtain the entanglement
between these two qubits using the formula for entanglement of
formation by Wootters \cite{EntanglementFormation,Entanglement-2qubits}. We estimate
values of phase parameters that give the maximum entanglement.

First, we consider the following pure state with only one global
phase parameter $\theta$,
\begin{equation}
|\psi_{ABC}\rangle
=
\frac{1}{2\sqrt{2}}
(e^{i\theta}|000\rangle
+|001\rangle+\cdots+|111\rangle).
\label{3qubit-1phase-parameter-pure-state1}
\end{equation}
Defining $\rho_{BC} =
\mbox{tr}_{A}|\psi_{ABC}\rangle\langle\psi_{ABC}|$, we obtain
$\rho_{AB}=\rho_{BC}=\rho_{CA}$ and we get
$E(\rho_{AB})=E(\rho_{BC})=E(\rho_{CA})$. If we decided to apply
the phase factor $e^{i\theta}$ to $|001\rangle$, instead of
$|000\rangle$, and calculated the entanglement between any two
qubits, we would obtain the same amount of entanglement as before.
To understand this, we apply $\mbox{\bf $I$}^{(A)} \otimes
\mbox{\bf $I$}^{(B)} \otimes \sigma_{x}^{(C)}$ to
Eq.~(\ref{3qubit-1phase-parameter-pure-state1}), and we obtain
\begin{eqnarray}
\lefteqn{|\psi'_{ABC}\rangle} \nonumber \\
&=& (\mbox{\bf $I$}^{(A)} \otimes \mbox{\bf
$I$}^{(B)} \otimes \sigma_{x}^{(C)})
|\psi\rangle_{ABC} \nonumber \\
&=&\frac{1}{2\sqrt{2}}
(|000\rangle
+e^{i\theta}|001\rangle
+|010\rangle+\cdots+|111\rangle).
\end{eqnarray}
In general, due to local convertibility, applying a phase factor
$e^{i\theta}$ to any of the kets $|x\rangle$ ($\forall x\in
\{0,1\}^{3}$) is equivalent in terms of entanglement as long as it
is the {\em only} phase which is applied.

Here, we evaluate the entanglement between the qubits $B$ and $C$
for the state given by
Eq.~(\ref{3qubit-1phase-parameter-pure-state1}). The reduced
density matrix for $\rho_{BC}$ for the qubits $B$ and $C$ is
given by
\begin{equation}
\rho_{BC}
=\frac{1}{8}
\left [
\begin{array}{cccc}
2 & 1+\tau & 1+\tau & 1+\tau \\
1+\bar{\tau} & 2 & 2 & 2 \\
1+\bar{\tau} & 2 & 2 & 2 \\
1+\bar{\tau} & 2 & 2 & 2 \\
\end{array}
\right ],
\label{density-matrix-BC-1para-pure}
\end{equation}
where $\tau=e^{i\theta}$. Before computing the entanglement, we
have to compute another density matrix $\tilde{\rho}$ from $\rho$
following the prescription given in
Ref.\cite{Entanglement-2qubits}. We get
\begin{equation}
\rho_{BC}\tilde{\rho}_{BC}
=
\frac{1}{64}
\left[
\begin{array}{cccc}
X & -X & -X & Z \\
Y & -Y & -Y & X \\
Y & -Y & -Y & X \\
Y & -Y & -Y & X
\end{array}
\right ],
\end{equation}
where
\begin{equation}
\begin{array}{rclcrcl}
X&=&-4\tau+|1+\tau|^{2},
&\quad&
Y&=&4(-1+\bar{\tau}), \\
Z&=&2(1-\tau^{2}).&&&&
\end{array}
\end{equation}

Defining an eigenvalue of $\rho_{BC}\tilde{\rho}_{BC}$ as $(\lambda/64)$,
we can write an equation for $\lambda$ as
\begin{equation}
\mbox{det}
|\rho_{BC}\tilde{\rho}_{BC}-\frac{\lambda}{64}\mbox{\bf $I$}| =0,
\end{equation}
and finally we obtain the following equation,
\begin{equation}
\lambda^{2}[\lambda^{2}+2\lambda(Y-X)+X^{2}-YZ]=0.
\end{equation}
Solutions of this equation are $\lambda=0$ for a double root
and
\begin{equation}
\lambda_{\pm}=2(\sqrt{2}\pm 1)^{2}(1-\cos \theta)(\geq 0).
\end{equation}

\begin{figure}[htbp]
\begin{center}
\leavevmode
\epsfxsize=8cm
\epsfbox{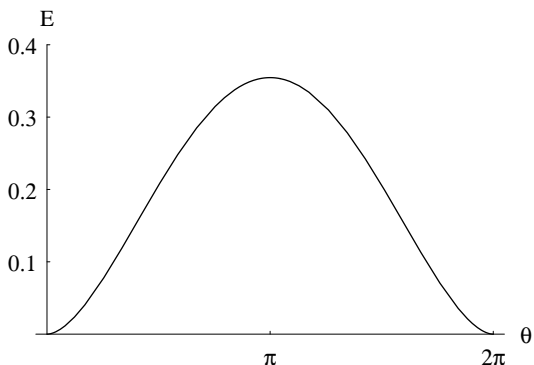}
\caption{The
entanglement $E$ against phase parameter $\theta$ for
Eq.~(\ref{density-matrix-BC-1para-pure}).}
\label{figure1}
\end{center}
\end{figure}

Therefore, because of $\lambda_{+}\geq\lambda_{-}$, the
concurrence \cite{Entanglement-2qubits} is given by
\begin{equation}
C=\frac{1}{8}(\sqrt{\lambda_{+}}-\sqrt{\lambda_{-}})
=\frac{1}{2\sqrt{2}}\sqrt{1-\cos\theta}(\geq 0),
\label{Concurrence-1para-pure}
\end{equation}
and entanglement can be written as ${\cal E}(C)=H(p)$,
where
\begin{equation}
p=\frac{1}{2}[1+\sqrt{1-\frac{1}{8}(1-\cos\theta)}].
\label{1para-eq-p-theta}
\end{equation}
From Eqs.~(\ref{Concurrence-1para-pure}) and
(\ref{1para-eq-p-theta}),
we find
\begin{equation}
0\leq C\leq\frac{1}{2},
\quad\quad
\frac{1}{2}(1+\frac{\sqrt{3}}{2})
\leq p \leq 1,
\label{C-p-ranges}
\end{equation}
where $p$ gets maximum at $\theta=0$ ($C=0$) and gets minimum at
$\theta=\pi$ ($C=1/2$).
(In this range, $H(p)$ decreases monotonously.)
$H(p)$ gets the maximum value of
$H((1/2)[1+(\sqrt{3}/2)])\simeq 0.36$ at $\theta=\pi$ and gets
the minimum one of $H(1)=0$ at $\theta=0$. In Fig.~\ref{figure1},
we show a variation of entanglement $E$ as a function of $\theta$.

The physical reason for the entanglement peaking at $\theta=\pi$
can be understood if $|\psi_{ABC}\rangle$ is rewritten in the
following manner
\begin{eqnarray}
|\psi_{ABC}\rangle
&\propto&
|0\rangle_A \otimes
(e^{i\theta}|00\rangle+|01\rangle+|10\rangle+|11\rangle)_{BC}\nonumber \\
&&
+|1\rangle_A \otimes
(|00\rangle+|01\rangle+|10\rangle+|11\rangle)_{BC}.
\end{eqnarray}
The state $\rho_{BC}$ is essentially a mixture of the state
$e^{i\theta}|00\rangle+|01\rangle+|10\rangle+|11\rangle$, which
is maximally entangled for $\theta=\pi$, and
$|00\rangle+|01\rangle+|10\rangle+|11\rangle$, which is always
disentangled.
(If we apply Hadamard transformation to the first qubit of the Bell-singlet,
we obtain
$-|00\rangle+|01\rangle+|10\rangle+|11\rangle$.)
Hence it is only expected that the entanglement of
the mixture will be maximum at $\theta=\pi$. It is also clear
that the entanglement can never be maximal in magnitude because an
entangled and a disentangled state are always mixed in equal
proportions in $\rho_{BC}$.

\begin{figure}[htbp]
\begin{center}
\leavevmode
\epsfxsize=8cm
\epsfbox{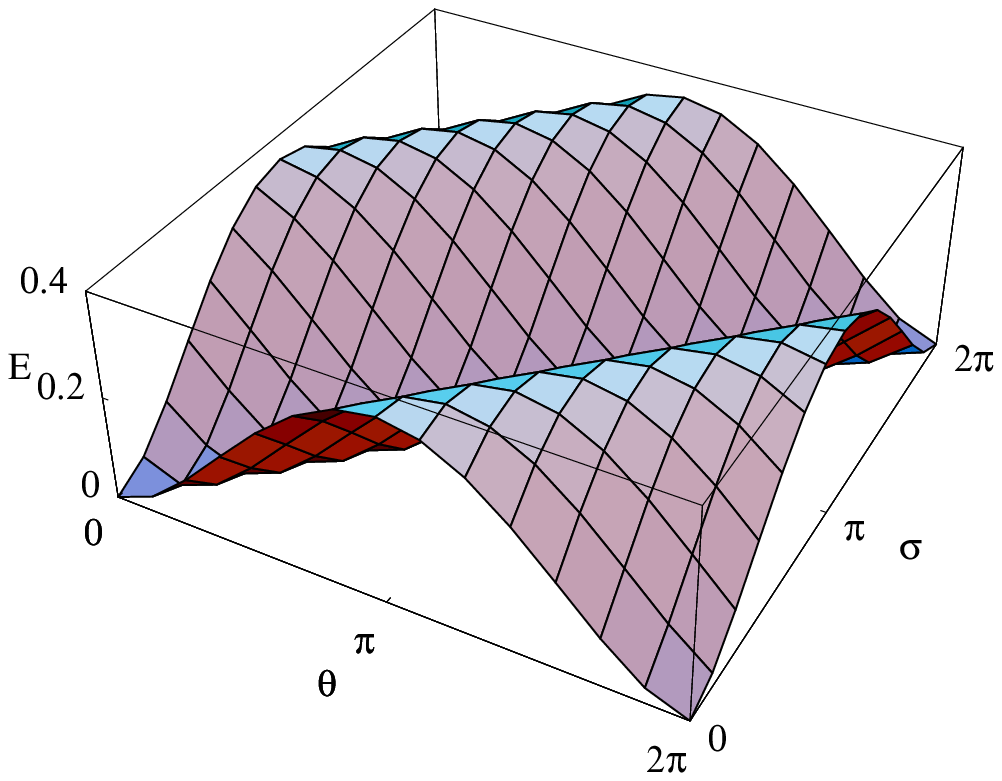}
\caption{The
entanglement $E$ against phase parameters $\theta$ and $\sigma$
for Eq.~(\ref{2qubit-DensityMatrix-2parameter1}).}
\label{figure2}
\end{center}
\end{figure}

Next, we consider pure states
with two phase parameters $\theta$ and $\sigma$. For example,
consider the following state,
\begin{eqnarray}
\lefteqn{|\psi_{ABC}\rangle}\nonumber \\
&=&
\frac{1}{2\sqrt{2}}
(e^{i\theta}|000\rangle
+e^{i\sigma}|001\rangle
+|010\rangle+\cdots+|111\rangle),
\label{3qubit-2phase-parameter-pure-state1}
\end{eqnarray}
we trace out the qubit $A$ and get
\begin{eqnarray}
\rho_{BC}
&=&
\frac{1}{8} {[} (e^{i\theta}|00\rangle
+e^{i\sigma}|01\rangle +|10\rangle+|11\rangle) \nonumber \\
&& \quad \times
(e^{-i\theta}\langle 00|+e^{-i\sigma}\langle 01|
+\langle 10|+\langle 11|) \nonumber \\
&&
+(|00\rangle+|01\rangle +|10\rangle+|11\rangle) \nonumber \\
&& \quad \times
(\langle 00|+\langle 01| +\langle 10|+\langle 11|){]} \nonumber \\
&=&
\frac{1}{8}
\left [
\begin{array}{cccc}
2 & \bar{\zeta}\tau+1 & \tau+1 & \tau+1 \\
\zeta\bar{\tau}+1 & 2 & \zeta+1 & \zeta+1 \\
\bar{\tau}+1 & \bar{\zeta}+1 & 2 & 2 \\
\bar{\tau}+1 & \bar{\zeta}+1 & 2 & 2 \\
\end{array}
\right ],
\label{2qubit-DensityMatrix-2parameter1}
\end{eqnarray}
where
$\tau=e^{i\theta}$ and $\zeta=e^{i\sigma}$.
Writing an eigenvalue of $\rho_{BC}\tilde{\rho_{BC}}$
as $(\lambda/64)$, we obtain $\lambda=0$ and
\begin{equation}
\lambda_{\pm}
=
2(\sqrt{2}\pm 1)^{2}
[1-\cos(\theta-\sigma)](\geq 0),
\label{Eigenvalues-rhorho-2para1-pure}
\end{equation}
Hence, the concurrence is
\begin{equation}
C=\frac{1}{2\sqrt{2}}
\sqrt{1-\cos(\theta-\sigma)}
(\geq 0),
\label{Concurrence-2para1-pure}
\end{equation}
and the entanglement can be written as
${\cal E}(C)=H(p)$,
where
\begin{equation}
p=\frac{1}{2}\{1+\sqrt{1-\frac{1}{8}[1-\cos(\theta-\sigma)]}\}.
\label{2para-eq-p-theta-sigma-1}
\end{equation}
From Eqs.~(\ref{Concurrence-2para1-pure})
and (\ref{2para-eq-p-theta-sigma-1}),
we find that $C$ and $p$ can take values
in the ranges of Eq.~(\ref{C-p-ranges}).
Because $p$ gets maximum at $\theta=\sigma$ ($C=0$)
and gets minimum at $\theta=\sigma\pm\pi$ ($C=1/2$),
$H(p)$ gets the maximum value at $\theta=\sigma\pm\pi$
and gets the minimum one at $\theta=\sigma$.
In Fig.~\ref{figure2},
we show a variation of entanglement $E$
as a function of $\theta$ and $\sigma$.

Again, in this case it is easy to see why the entanglement is
minimum for $\theta=\sigma$. The whole state can be rewritten as
\begin{eqnarray}
\lefteqn{|\psi_{ABC}\rangle} \nonumber \\
&\propto&
|0\rangle_A \otimes
{[}e^{i\theta}|0\rangle(|0\rangle+e^{i(\sigma-\theta)}|1\rangle)
+|1\rangle(|0\rangle+|1\rangle){]}_{BC}\nonumber \\
&&
+|1\rangle_A \otimes
(|0\rangle+|1\rangle)_{B}(|0\rangle+|1\rangle)_{C}.
\end{eqnarray}
This makes it clear that the state $\rho_{BC}$ is a mixture of the
state
${[}e^{i\theta}|0\rangle(|0\rangle+e^{i(\sigma-\theta)}|1\rangle)
+|1\rangle(|0\rangle+|1\rangle){]}_{BC}$,
which is entangled for $\sigma \neq \theta$, and the always
disentangled state
$(|0\rangle+|1\rangle)_{B}(|0\rangle+|1\rangle)_{C}$. The
entanglement of $\rho_{BC}$ will thus depend entirely on the
entanglement of
${[}e^{i\theta}|0\rangle(|0\rangle+e^{i(\sigma-\theta)}|1\rangle)
+|1\rangle(|0\rangle+|1\rangle){]}_{BC}$,
whose entanglement will be zero when $\theta=\sigma$ and maximum when $\theta-\sigma=\pi$.

The entanglement between the qubits $B$ and $C$ will depend on the choice of the
two kets from the set $\{|x\rangle : x \in \{0,1\}^{3}\}$
to which we decide to apply the global phases
($e^{i\theta}$ and $e^{i\sigma}$) (It is different from the
one-parameter case of
Eq.~(\ref{3qubit-1phase-parameter-pure-state1})). Imagine that we
had applied the phases to $|000\rangle$ and $|011\rangle$. Then
the reduced density matrix for $\rho^{'}_{BC}$ would be
\begin{eqnarray}
\rho^{'}_{BC}
&=&
\frac{1}{8} {[}
(e^{i\theta}|00\rangle+|01\rangle
+|10\rangle +e^{i\sigma}|11\rangle) \nonumber \\
&&\quad\times
(e^{-i\theta}\langle 00|+\langle 01|
+\langle 10|+e^{-i\sigma}\langle 11|) \nonumber \\
&&
+(|00\rangle+|01\rangle +|10\rangle+ |11\rangle) \nonumber \\
&&\quad\times
(\langle 00|+\langle 01| +\langle 10|+\langle 11|) {]}.
\label{2qubit-DensityMatrix-2parameter2}
\end{eqnarray}
Because we cannot transform the density matrix $\rho$ of
Eq.~(\ref{2qubit-DensityMatrix-2parameter2}) to that of
Eq.~(\ref{2qubit-DensityMatrix-2parameter1}) by local unitary
transformations $U^{(A)}\otimes U^{(B)}\otimes U^{(C)}$, the
entanglement of Eq.~(\ref{2qubit-DensityMatrix-2parameter1}) need
not be equal to that of
Eq.~(\ref{2qubit-DensityMatrix-2parameter2}) in general.

\begin{figure}[htbp]
\begin{center}
\leavevmode
\epsfxsize=8cm
\epsfbox{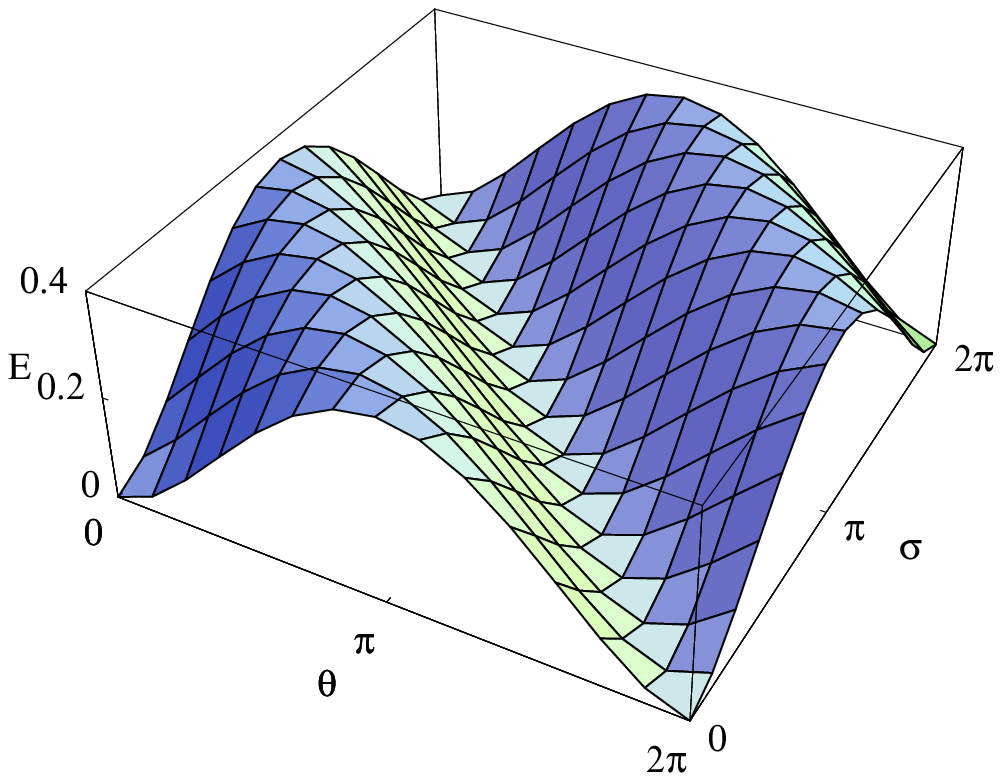}
\caption{The
entanglement $E$ against phase parameters $\theta$ and $\sigma$
for Eq.~(\ref{2qubit-DensityMatrix-2parameter2}).}
\label{figure3}
\end{center}
\end{figure}

Writing $\rho^{'}_{BC}$ as
\begin{equation}
\rho^{'}_{BC}
=
\frac{1}{8}
\left [
\begin{array}{cccc}
2 & \tau+1 & \tau+1 & \tau\bar{\zeta}+1 \\
\bar{\tau}+1 & 2 & 2 & \bar{\zeta}+1 \\
\bar{\tau}+1 & 2 & 2 & \bar{\zeta}+1 \\
\bar{\tau}\zeta+1 & \zeta+1 & \zeta+1 & 2 \\
\end{array}
\right ],
\end{equation}
and an eigenvalue of $\rho^{'}_{BC}\tilde{\rho}^{'}_{BC}$ as $(\lambda/64)$,
we obtain $\lambda=0$
and
\begin{equation}
\lambda_{\pm}
=
2\{3r+2ts\pm 2[2r(r+ts)]^{1/2}\},
\label{Eigenvalues-rts}
\end{equation}
where
$t=1-\cos\theta$,
$s=1-\cos\sigma$,
and $r=1-\cos(\theta+\sigma)$.
The concurrence is given by
$C=(1/8)(\sqrt{\lambda_{+}}-\sqrt{\lambda_{-}})$ and $0\leq C\leq
1/2$. At $\theta+\sigma=0$ $(\mbox{mod $2\pi$})$, $C=0$ and the
entanglement $E$ gets minimum. At $\theta+\sigma=\pi$ $(\mbox{mod
$2\pi$})$, $C=1/2$ and $E$ gets maximum. In Fig.~\ref{figure3},
we show a variation of entanglement $E$ as a function of $\theta$
and $\sigma$. As in the previous cases, the entanglement is
entirely due to the entanglement of the first part
$e^{i\theta}|00\rangle+|01\rangle +|10\rangle
+e^{i\sigma}|11\rangle$ of the density matrix $\rho^{'}_{BC}$.

\begin{figure}[htbp]
\begin{center}
\leavevmode
\epsfxsize=8cm
\epsfbox{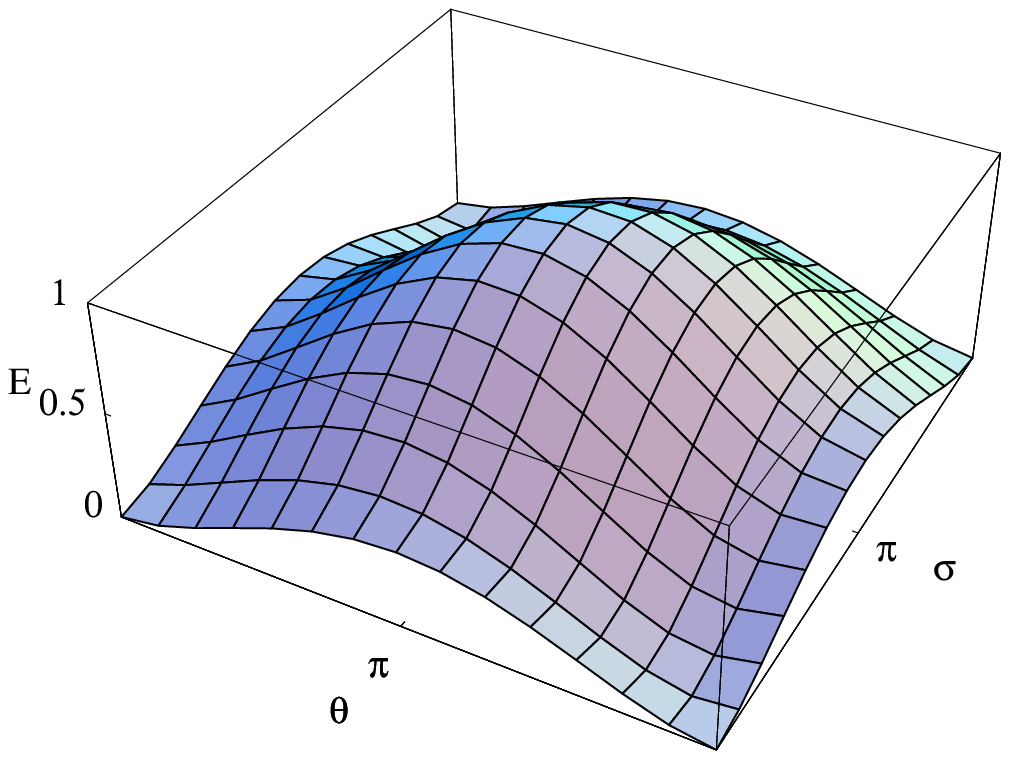}
\caption{The entanglement
$E$ against phase parameters $\theta$ and $\sigma$ for
Eq.~(\ref{2qubit-rho-AB-2parameter-a}).}
\label{2paraEntA1}
\end{center}
\end{figure}

Note that in both the cases of Eqs.~(\ref{2qubit-DensityMatrix-2parameter1})
and (\ref{2qubit-DensityMatrix-2parameter2}) maximal entanglement
between $B$ and $C$ can never be reached by varying $\theta$ and
$\sigma$. However, one could get maximal entanglement if one
applied the two phase parameters to two different global states.
This is equivalent to applying same sets of phases as before, but
examining the entanglement between the pair of qubits $A$ and $C$
or $A$ and $B$. Let us consider the three-qubit pure state
of Eq.~(\ref{3qubit-2phase-parameter-pure-state1}) again
and trace out the qubit $C$ (in contrast to $A$) to get
\begin{eqnarray}
\rho_{AB}
&=&
\frac{1}{8} {[} (e^{i\theta}|00\rangle+|01\rangle
+|10\rangle+|11\rangle) \nonumber \\
&&\quad \times
(e^{-i\theta}\langle 00|+\langle 01|
+\langle 10|+\langle 11|) \nonumber \\
&&
+(e^{i\sigma}|00\rangle+|01\rangle
+|10\rangle+|11\rangle) \nonumber \\
&&\quad \times
(e^{i\sigma}\langle 00|+\langle 01|
+\langle 10|+\langle 11|)
{]} \nonumber \\
&=& \frac{1}{8} \left [
\begin{array}{cccc}
2 & \zeta+\tau & \zeta+\tau & \zeta+\tau \\
\bar{\zeta}+\bar{\tau} & 2 & 2 & 2 \\
\bar{\zeta}+\bar{\tau} & 2 & 2 & 2 \\
\bar{\zeta}+\bar{\tau} & 2 & 2 & 2 \\
\end{array}
\right ].
\label{2qubit-rho-AB-2parameter-a}
\end{eqnarray}
If we write an
eigenvalue of $\rho_{AB}\tilde{\rho_{AB}}$ as $(\lambda/64)$, we
obtain $\lambda=0$ and
\begin{equation}
\lambda_{\pm}=
2{[}
4(t+s)-u \pm 2\{2(t+s)
{[}2(t+s)-u{]}\}^{1/2}{]},
\end{equation}
where
$t$ and $s$ are defined before
and $u=1-\cos(\theta-\sigma)$.
In Fig.~\ref{2paraEntA1}, we show a variation of entanglement
of $\rho_{AB}$ as a function of $\theta$ and $\sigma$.

\begin{figure}[htbp]
\begin{center}
\leavevmode
\epsfxsize=8cm
\epsfbox{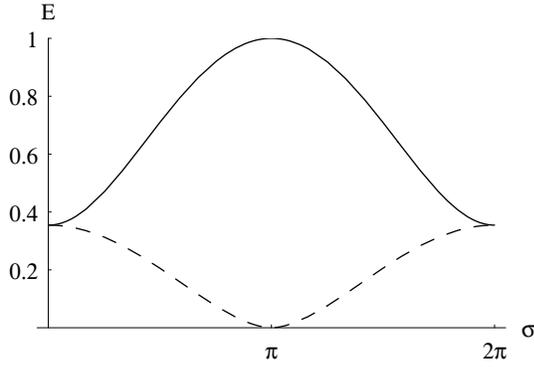}
\caption{The entanglement $E$ of $\rho_{AB}$ for
Eq.~(\ref{2qubit-rho-AB-2parameter-a})
and $\rho_{BC}$ for
Eq.~(\ref{2qubit-DensityMatrix-2parameter1})
against $\sigma$ with fixed $\theta(=\pi)$. A solid line
represents $E$ of $\rho_{AB}$ and a dashed line represents $E$ of
$\rho_{BC}$.}
\label{2paraEntA2}
\end{center}
\end{figure}

We now compare the entanglement of $\rho_{AB}$
for Eq.~(\ref{2qubit-rho-AB-2parameter-a})
and $\rho_{BC}$
for Eq.~(\ref{2qubit-DensityMatrix-2parameter1})
with fixed $\theta$. In Fig.~\ref{2paraEntA2}, we show the
variation of entanglement of $\rho_{AB}$ and $\rho_{BC}$ with
$\theta=\pi$. From Fig.~\ref{2paraEntA2}, we notice the following
facts. When the entanglement $E$ of $\rho_{BC}$ decreases, $E$ of
$\rho_{AB}$ increases. $\rho_{AB}$ becomes the maximally
entangled state at $\theta=\sigma=\pi$. To understand this, we
rewrite $\rho_{AB}$ with $\theta=\pi$ as follows
\begin{eqnarray}
\rho_{AB}
&=&
\frac{1}{4} {[}
(|0\rangle|-\rangle+|1\rangle|+\rangle)
(\langle 0|\langle-|+\langle 1|\langle +|) \nonumber \\
&&
+
(|0\rangle|\phi_{\sigma}\rangle+|1\rangle|+\rangle)
(\langle 0|\langle \phi_{\sigma}|+\langle 1|\langle +|){]},
\label{rho-A-fixed-theta-pi}
\end{eqnarray}
where
\begin{eqnarray}
|-\rangle&=&(1/\sqrt{2})(-|0\rangle+|1\rangle), \nonumber \\
|+\rangle&=&(1/\sqrt{2})(|0\rangle+|1\rangle), \nonumber \\
|\phi_{\sigma}\rangle&=&
(1/\sqrt{2})(e^{i\sigma}|0\rangle+|1\rangle).
\end{eqnarray}
Note that $(|0\rangle|-\rangle+|1\rangle|+\rangle)$ is the
maximally entangled state and the phase parameter $\sigma$
controls the entanglement of the second term in
Eq.~(\ref{rho-A-fixed-theta-pi}). As
$|\phi_{\sigma=\pi}\rangle=|-\rangle$,  $A$ and $B$ are maximally
entangled in the state $(|0\rangle|-\rangle+|1\rangle|+\rangle)$
(with $C$ being completely disentangled from them) for
$\sigma=\pi$.

\section{Entanglement between two qubits for three-qubit mixed states}
\label{Entanglement-on-3qubit-mixed-state}

In previous sections, we have studied the entanglement between
two qubits on pure states with phase factors. The pure state of
Eq.~(\ref{3qubit-1phase-parameter-pure-state1}) is prepared by
taking a three-qubit states $(|+\rangle\langle +|)^{\otimes 3}$,
and giving a phase $e^{i\theta}$ on the ket vector $|000\rangle$.
(In this section, we will often use the basis
$\{|\pm\rangle=(1/\sqrt{2})(|0\rangle\pm|1\rangle)\}$.)

Here, instead of the pure state $(|+\rangle\langle +|)^{\otimes 3}$,
we take a mixed state,
\begin{equation}
[(1-q)|+\rangle\langle +|+q|-\rangle\langle -|]^{\otimes 3},
\label{MixedState-NoPhasePara}
\end{equation}
where $0\leq q \leq 1/2$. Then, we consider the application of a
single phase factor as follows
\begin{equation}
|000\rangle \rightarrow e^{i\theta}|000\rangle.
\end{equation}
Tracing out any qubit out of the three qubits,
we obtain the density matrix in the form of
\begin{equation}
\rho=\frac{1}{4}
\left [
\begin{array}{cccc}
1 & (1+\tau)\alpha & (1+\tau)\alpha & 2(1+\tau)\alpha^{2} \\
(1+\bar{\tau})\alpha & 1 & 4\alpha^{2} & 2\alpha \\
(1+\bar{\tau})\alpha & 4\alpha^{2} & 1 & 2\alpha \\
2(1+\bar{\tau})\alpha^{2} & 2\alpha & 2\alpha & 1 \\
\end{array}
\right ],
\label{density-operator-mixed-1para}
\end{equation}
where $\tau=e^{i\theta}$ and $\alpha=(1/2)-q$. Now we proceed to
derive the entanglement $E(\rho)$ as a function of $\theta$ and
$q$.

We already know that entanglement takes the maximum value at
$\theta=\pi$ when we fix $q=0$. The interesting question is
whether that peak of entanglement remains in the same place for a
nonzero $q$. Before evaluating $E(\rho)$ explicitly, we show that
it gets a local stationary value at $\theta=\pi$ for arbitrary
fixed $p$ ($0\leq \forall q \leq 1/2$). (It remains stationary
locally along $\theta$-axis at any fixed $q$.)

We first show that an infinitesimal variation of $\theta$ from
$\theta=\pi$ does not affect an equation of eigenvalues of
$\rho\tilde{\rho}$. The equation of eigenvalues of
$\rho\tilde{\rho}$ with $\theta=\pi+\delta$ and $|\delta|\ll 1$
is given by
\begin{eqnarray}
\lefteqn{\mbox{det}|\rho\tilde{\rho} -\lambda\mbox{\bf $I$}|
{\:\rule[-1ex]{.05em}{3ex}\,}_{\theta=\pi+\delta}} \nonumber \\
&=&
\mbox{det}|\rho\tilde{\rho} -\lambda\mbox{\bf $I$}|
{\:\rule[-1ex]{.05em}{3ex}\,}_{\theta=\pi}
+ \delta
\frac{\partial}{\partial\theta} [\mbox{det}|\rho\tilde{\rho}
-\lambda\mbox{\bf $I$}|]
{\:\rule[-1ex]{.05em}{3ex}\,}_{\theta=\pi} + O(\delta^{2})
\nonumber \\
&=& 0.
\end{eqnarray}
Hence, if $\partial_{\theta} [\mbox{det}|\rho\tilde{\rho}
-\lambda\mbox{\bf $I$}|]
{\:\rule[-.5ex]{.05em}{2.5ex}\,}_{\theta=\pi}=0$, the equation is
not affected by $\delta$ and the eigenvalues of
$\rho\tilde{\rho}$ get stationary around a neighborhood of
$\theta=\pi$ for fixed $q$.

Writing
\begin{equation}
\rho\tilde{\rho}-\lambda\mbox{\bf $I$} = \left [
\begin{array}{cccc}
X+L & V & V & W \\
Y & -Z+L & -Z & -V \\
Y & -Z & -Z+L & -V \\
Z & -Y & -Y & X+L \\
\end{array}
\right ],
\end{equation}
where
\begin{eqnarray}
X&=&-(1/16)(1-\bar{\tau})\alpha^{2}
[(1-\tau)\alpha^{2}+\tau], \nonumber \\
Y&=&-(1/16)(1-\bar{\tau})\alpha
(1+4\alpha^{2}), \nonumber \\
Z&=&-(1/4)(1-\bar{\tau})\alpha^{2}, \nonumber \\
V&=&(1/16)(1-\bar{\tau})\alpha
[2\alpha^{2}+\tau(1+2\alpha^{2})], \nonumber \\
W&=&(1/8)(1-\tau^{2})\alpha^{2}, \nonumber \\
L&=&-\lambda
+(1/16)(1+2\alpha)^{2}(1-2\alpha)^{2}, \nonumber \\
\end{eqnarray}
we can obtain the following result
with some calculations\cite{Matrix-textbook},
\begin{eqnarray}
\lefteqn{\frac{\partial}{\partial\theta}
[\mbox{det}|\rho\tilde{\rho} -\lambda\mbox{\bf $I$}|]
{\:\rule[-1ex]{.05em}{3ex}\,}_{\theta=\pi}} \nonumber \\
&=&[
\left |
\begin{array}{cccc}
\partial_{\theta}X & \partial_{\theta}V &
\partial_{\theta}V & \partial_{\theta}W \\
Y & -Z+L & -Z & -V \\
Y & -Z & -Z+L & -V \\
Z & -Y & -Y & X+L \\
\end{array}
\right | \nonumber \\
&&
+2
\left |
\begin{array}{cccc}
X+L & V & V & W \\
\partial_{\theta}Y & -\partial_{\theta}Z &
-\partial_{\theta}Z & -\partial_{\theta}V \\
Y & -Z & -Z+L & -V \\
Z & -Y & -Y & X+L \\
\end{array}
\right | \nonumber \\
&&
+
\left |
\begin{array}{cccc}
X+L & V & V & W \\
Y & -Z+L & -Z & -V \\
Y & -Z & -Z+L & -V \\
\partial_{\theta}Z & -\partial_{\theta}Y &
-\partial_{\theta}Y & \partial_{\theta}X \\
\end{array}
\right |
]
{\:\rule[-1ex]{.05em}{3ex}\,}_{\theta=\pi} \nonumber \\
&=&0.
\end{eqnarray}
Therefore,
$E(\rho)$ remains stationary at $\theta=\pi$ for any fixed $q$
and we can expect that it gets maximum there along $\theta$-axis.

\begin{figure}[htbp]
\begin{center}
\leavevmode
\epsfxsize=8cm
\epsfbox{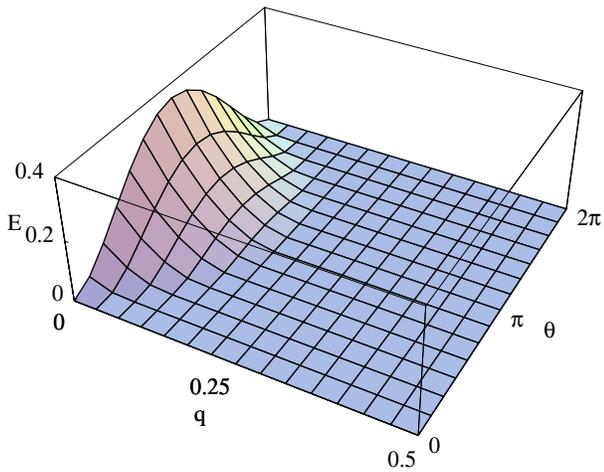}
\caption{The
entanglement $E$ against phase parameter $\theta$ and probability
$q$ for the mixed state of
Eq.~(\ref{density-operator-mixed-1para}).}
\label{figure4}
\end{center}
\end{figure}

\begin{figure}[htbp]
\begin{center}
\leavevmode
\epsfxsize=8cm
\epsfbox{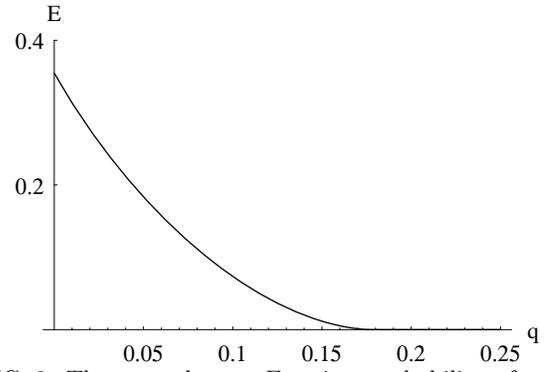}
\caption{The
entanglement $E$ against probability $q$ for $\theta=\pi$ for the
mixed state of Eq.~(\ref{density-operator-mixed-1para}).}
\label{figure5}
\end{center}
\end{figure}

By numerical calculations, we get Fig.~\ref{figure4}. It is clear
from this figure that the basic behaviour of entanglement with
variation of a single phase parameter $\theta$ does not change for
a mixed initial state and it is still maximum at $\theta=\pi$.
Fig.~\ref{figure5} shows variation of $E$ as a function of $q$
for $\theta=\pi$. This figure illustrates that the entanglement is
lost rapidly as $q$ gets larger. This is also an expected result:
the more mixed the initial state is, the harder it is to entangle
it by global phase functions.

\section{Necessity of entanglement for exponential speedup in Deutsch-Jozsa algorithm}
We now present an application of our results on entangling by
global phases to the question of necessity of entanglement in
quantum computation. In the Deutsch-Jozsa algorithm, the
following state appears\cite{dj},
\begin{equation}
|\Psi\rangle
=2^{-n/2}
\sum_{j\in\{0,1\}^{n}} e^{if(j)}|j\rangle,
\end{equation}
where $0\leq f(j)\leq 2\pi$ for $\forall j$. If $f(j)$ is
constant for $\forall j$, $|\Psi\rangle$ is a uniform
superposition, and we get $|0\cdots 0\rangle$ by applying the
quantum Fourier transformation (QFT) to $|\Psi\rangle$. On the
other hand, if $\{f(j)\}$ takes on values $0$ or $\pi$ randomly
but in a balanced manner (i.e. equal occurrences of $0$ and
$\pi$), $|\Psi\rangle$ is orthogonal to the uniform superposition
and we get a state orthogonal to $|0\cdots 0\rangle$ after QFT.
Therefore, we can investigate whether $f$ is constant or balanced
by a single application of the global phase function using a
quantum computer. On the other hand, in the worst case scenario
using a classical algorithm, one may have to evaluate this
function for at least half the number of possible arguments $j$.
This implies $2^n/2$ (exponential) function evaluations. This is
why Deutsch-Jozsa algorithm is regarded as having an exponential
advantage over its classical counterpart.

To see that entanglement is necessary for the exponential
advantage of this algorithm, consider the following scenario. It
is given that the global phase functions, apart from being
constant or balanced and taking values $0$ or $\pi$, are also
restricted in such a manner that they never produce an entangled
state in the course of the entire computation. This implies
(according to the conditions obtained in Section
\ref{Disentanglement}),
\begin{equation}
f(j)=\vec{\theta}\cdot\vec{j}+\theta_{0} \quad\quad (\mbox{mod
$2\pi$}). \label{product-state-condition}
\end{equation}
If we know beforehand that $f$ can be written as
Eq.~(\ref{product-state-condition}), we can estimate $f$
completely with $O(n)$ steps of classical algorithm, even in the
worst case. We supply $(0\cdots 0)$ and strings where only one
digit is $1$ and the others are $0$, $(10\cdots 0)$, $\cdots$,
$(0\cdots 01)$, as $j$ of inputs for $f$, and we get $\theta_{0}$
and $\bar{\theta}$ as outputs. Hence, when we restrict the
possible set of functions to those which are {\em non-entanglement
producing}, a polynomial time classical algorithm exists. In other
words, there is only a polynomial advantage of quantum computation
over classical computation.  To make the quantum algorithm have an
exponential advantage over its classical counterpart, we must
remove the restriction of Eq.~(\ref{product-state-condition}) on
the global phase functions $f(j)$, which implies that entanglement
cannot be prevented from arising any more during the course of the
quantum computation. As no entanglement implies only polynomial
advantage, to get exponential advantage, entanglement is
necessary.

\section{Conclusions}
In this paper, we have investigated the generation of entanglement
through global phase functions. We have obtained necessary and
sufficient conditions for the application of global phases to the
pure product state
$|0 \cdots 00\rangle+|0 \cdots 01\rangle+ \cdots +|1 \cdots 11\rangle$ to
result in entanglement. We have then investigated the amount of
two qubit entanglement that can be generated in three-qubit pure
states when only one or two of the global phase parameters are
nonzero. An interesting, though potentially difficult, future
direction will be in the investigation of the quantity of
entanglement when all phase parameters are present for an
arbitrary number of qubits. While we have obtained the conditions
for {\em presence} or {\em absence} of entanglement in the general
case, it would be interesting to classify functions according to
the {\em degree} of entanglement they can generate.  We have also
examined entanglement generation through a single global phase
parameter for mixed initial states. The general problem of finding
necessary and sufficient conditions for entanglement by global
phases for mixed states remains open. One could expect
counterintuitive results in that case as the same global phase
function might entangle one pure component and disentangle another
pure component of a mixture of two pure states. Finally, we have
applied our conditions to prove the necessity of entanglement in
the Deutsch-Jozsa algorithm for the algorithm to have an
exponential advantage over its classical counterpart. It would be
interesting to apply similar techniques to the investigation of
the role of entanglement in other quantum algorithms.

\newpage

\begin{references}
%
\bibitem{bennett00}
C.~H.~Bennett and D.~P.~DiVincenzo, Nature {\bf 404}, 247 (2000).
%
\bibitem{Deu85}
D.~Deutsch, Proc. R. Soc. Lond. A {\bf 400}, 97 (1985).
%
\bibitem{dj}
D.~Deutsch and R.~Jozsa, Proc. R. Soc. Lond. A {\bf 439}, 553 (1992).
%
\bibitem{Shor}
P.~W.~Shor,
``Algorithms for Quantum Computation:
Discrete Logarithms and Factoring,''
{\it in Proc. 35th Ann. Symp.
on the Foundations of Computer Science}
(IEEE Computer Society, Los Alamitos, 1994),
pp. 124--134;
P.~W.~Shor,
SIAM J. Comput. {\bf 26}, 1484 (1997).
%
\bibitem{Grov}
L.~K.~Grover,
``A fast quantum mechanical algorithm for database search,''
{\it in Proc. 28th Ann. ACM Symp.
on the Theory of Computing}
(ACM Press, New York, 1996) pp. 212--219;
L.~K.~Grover,
Phys. Rev. Lett. {\bf 79}, 325 (1997).
%
\bibitem{jozsa}
R.~Jozsa, ``Entanglement and Quantum Computation,''
{\it The Geometric Universe:
Science, Geometry, and the Work of Roger Penrose,}
ed. S.~A.~Huggett {\it et al.}, pp. 369--379,
(Oxford University Press, 1998).
%
\bibitem{ej}
A.~Ekert and R.~Jozsa, Phil. Trans. R. Soc. A {\bf 356}, 1769 (1998).
%
\bibitem{lindpop}
N.~Linden and S.~Popescu,
``Good dynamics versus bad kinematics.
Is entanglement needed for quantum computation?'',
LANL eprint quant-ph/9906008.
%
\bibitem{mosca}
R.~Cleve, A.~Ekert, C.~Macchiavello and M.~Mosca,
Proc. R. Soc. Lond. A {\bf 454}, 339 (1998).
%
\bibitem{EntanglementFormation}
C.~H.~Bennett, D.~P.~DiVincenzo, J.~A.~Smolin
and W.~K.~Wootters,
Phys. Rev. A {\bf 54}, 3824 (1996);\\
C.~H.~Bennett and P.~W.~Shor,
IEEE Trans. on Information Theory {\bf 44}, 2724 (1998).
%
\bibitem{Entanglement-2qubits}
S.~Hill and W.~K.~Wootters,
Phys. Rev. Lett. {\bf 78}, 5022 (1997);
W.~K.~Wootters,
Phys. Rev. Lett. {\bf 80}, 2245 (1998).
%
\bibitem{Matrix-textbook}
S.~Barnett, {\it Matrices: Methods and Applications},\\
Chapt.~4
(Clarendon Press, Oxford, 1990).
\end{references}
\end{document}